\pdfoutput=1
\documentclass[aps,pra,reprint,amsmath,amssymb]{revtex4-2}
\usepackage{bm}
\usepackage{graphicx}
\usepackage[colorlinks]{hyperref}
\usepackage{mathrsfs}
\usepackage{times}

\begin{document}

\title{Krylov Spread Complexity of Quantum-Walks}

\author{Bhilahari Jeevanesan}
\email{Bhilahari.Jeevanesan@dlr.de}
\affiliation{Remote Sensing Technology Institute, German Aerospace Center DLR, 82234 Wessling, Germany}

\begin{abstract}
Given the recent advances in quantum technology, the complexity of quantum states is an important notion. The idea of the \emph{Krylov spread complexity} has come into focus recently with the goal of capturing this in a quantitative way. The present paper sheds new light on the Krylov complexity measure by exploring it in the context of continuous-time quantum-walks on graphs. A close relationship between Krylov spread complexity and the concept of \emph{limiting-distributions} for quantum-walks is established. Moreover, using a graph optimization algorithm, quantum-walk graphs are constructed that have vertex states with minimal and maximal (long-time average) Krylov $\bar{\mathcal{C}}$-complexity. This reveals an empirical upper bound for the $\bar{\mathcal{C}}$-complexity as a function of Hilbert space dimension and an exact lower bound.
\end{abstract}

\maketitle
\section{Introduction}
In the past years, the need to understand the complexity of quantum states and quantum operators has independently emerged in several subfields of physics. With the recent advances around noisy intermediate-scale quantum (NISQ) devices \cite{preskill2018quantum}, implementing quantum states as efficiently as possible has become a practical necessity. NISQ devices are severely limited by noise and short coherence times, thus the employable number of gates tends to be small. The minimum required number of gates to approximately prepare a given unitary is known as the \emph{gate complexity} and determines the feasibility of loading states on NISQ devices.

Independently, in the quantum gravity community, questions about quantum complexity have arisen in the context of understanding the interior of black holes through holographic dualities \cite{brown2018second, susskind2020three,stanford2014complexity}. This in turn, has rekindled interest in the beautiful quantum information work initiated by Nielsen and co-workers \cite{nielsen2005geometric, dowling2008geometry,nielsen2006optimal} on the connection between gate complexity and curved space geometries. By introducing a metric on the $\text{SU}(2^N)$ unitary group, the authors sketched out a program to numerically quantify the complexity of any given unitary $U$ via its geodesic distance to the identity operator. In practice, however, this approach has been difficult to carry out for many qubits and remains formidable even in the single-qubit case \cite{brown2019complexity}.

Consequently, a great deal of effort has been expended in studying alternative complexity notions such as the \emph{Krylov complexity} introduced in \cite{parker2019universal} for quantum operators, which has triggered a large wave of follow-up works \cite{rabinovici2022krylov, barbon2019evolution, rabinovici2021operator, muck2022krylov, caputa2022geometry,Bhattacharya:2022aa, hashimoto2023krylov,hornedal2022ultimate,dymarsky2021krylov, liu2023krylov,bhattacharjee2024operator,suchsland2023krylov, camargo2024spectral}. Further extending this notion, the idea of the \emph{spread-complexity} of a quantum state was introduced in \cite{balasubramanian2022quantum}. Interesting results have been obtained for this complexity measure in problems such as the classification of topological phases of matter \cite{caputa2022quantum,caputa2023spread}, detection of scar states in many-body systems \cite{bhattacharjee2022probing}, non-unitary quantum dynamics \cite{bhattacharya2024spread}, quantum billiards \cite{camargo2024spectral} and for saddle-dominated scrambling \cite{huh2024spread}.
A recent paper \cite{aguilar2024krylov} points out that Krylov complexity cannot equal the Nielsen complexity, since the former does not satisfy the fundamental axioms of distance measures. Another recent work \cite{PhysRevLett.132.160402} shows how a link between these two concepts can be established nevertheless  through a common matrix that appears in both definitions. Despite such works, the precise connection between Krylov spread complexity and gate complexity has remained elusive. Clearly, further investigations of both complexity measures are required.

\begin{figure}
\centering{}\includegraphics[width=0.9\columnwidth]{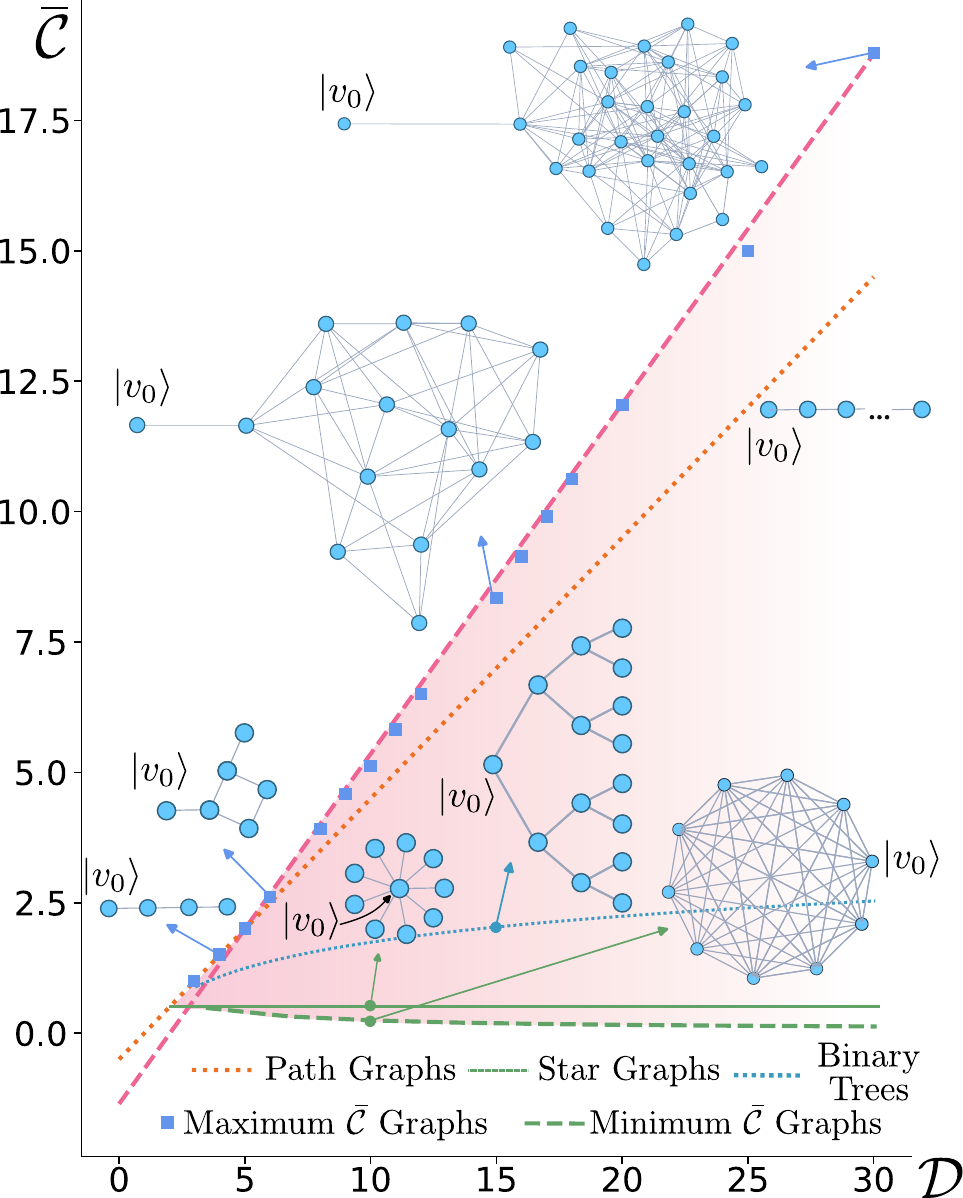}\caption{\label{FigMain}  The  Krylov $\bar{\mathcal C}$-complexity as a function of Hilbert space dimension $\mathcal D$ for various classes of graphs. The minimum $\bar{\mathcal C}$-complexity (dashed green) is attained by complete graphs. The data points (blue squares) show the largest possible $\bar{\mathcal C}$ values for a given $\mathcal D$ together with some of the corresponding graphs. The results were obtained by running a graph optimization algorithm for values up to $\mathcal D = 30$. All $\bar{\mathcal C}$-complexities are found to lie in the shaded wedge.}
\end{figure}

In the present work, we seek to shed new light on the concept of Krylov complexity by applying it to quantum walks. Quantum walks are of great interest since they constitute a framework for universal quantum computation, in fact as shown by Childs et al. \cite{childs2009universal, childs2013universal} any unitary operation on qubits can be realized by constructing a suitable unweighted graph. 

The first result of the present paper is the diagram in \mbox{Fig. \ref{FigMain}}, showing that the possible range of $\bar{\mathcal{C}}$ complexity, to be defined below, lies in the shaded wedge. The maximum possible complexity satisfies the asymptotic relation eq. \eqref{maxComplxty}. The second result ties the notion of Krylov $\bar{\mathcal{C}}$-complexity together with the well-established concept of \emph{limiting-distributions} $\chi$ for quantum walks. As a bonus, the Krylov approach allows us to compute the exact value of $\chi$ at the exit node of the glued binary tree Fig. \ref{GluedBinaryTree}.

\section{Krylov Spread Complexity and its Long-Time Average}
We work with the notion of Krylov spread-complexity introduced in \cite{balasubramanian2022quantum} and investigate quantum-walk Hamiltonians that give rise to states with minimal and maximal complexity. Informally speaking, the Krylov spread-complexity of a state $|\psi_0\rangle$ captures how compactly the state $|\psi(t)\rangle=\exp{(-i H t)}|\psi_0\rangle$ can be described. For very small $t$ the state $|\psi(t)\rangle$ can be described by $|\psi_0\rangle$ and $H|\psi_0\rangle$. As time progresses, more powers of the Hamiltonian $H$ are required, thus we consider the subspace that is spanned by $\{|\psi_0\rangle,H|\psi_0\rangle,H^2|\psi_0\rangle,\dots\}$. Performing a Gram-Schmidt orthonormalization on the states in this order, one obtains the Krylov-basis $\{K_n\}$. The spread-complexity of $|\psi_0\rangle$ in a $\mathcal D$-dimensional Hilbert space is then defined as
\begin{eqnarray}
\label{eq:cmplxDef}
\mathcal C(t)\equiv \sum_{n=0}^{\mathcal D -1} w_n|\langle K_n|\psi(t)\rangle|^2,
\end{eqnarray}
where ${w_n}$ is a sequence of positive increasing real numbers. Below we will set $w_n = n$. 
The intuition behind this definition is that states requiring the higher-indexed Krylov-basis vectors for their description are more complex, since their descriptions require higher powers $H^m |\psi_0\rangle$. It has been shown in \cite{balasubramanian2022quantum} that the Krylov-basis is optimal in the sense that there is a time interval $[0,T]$ for which the value of complexity as defined in eq. \eqref{eq:cmplxDef} cannot be decreased by using any other basis in its place. The Krylov basis satisfies the relation \cite{lanczos1950iteration}
\begin{eqnarray}
\label{eq:KrylovBasis}
H |K_n\rangle = a_n |K_n\rangle + b_{n} |K_{n-1}\rangle+b_{n+1}|K_{n+1}\rangle,
\end{eqnarray}
with complex numbers $a_n$, $b_n$. The basis vectors are defined by starting from  $|K_0\rangle = |\psi_0\rangle$ with $b_0=0$. The remaining $b_n$ are chosen such that each $|K_n\rangle$ is normalized and orthogonal to $|K_{n-1}\rangle$ and $|K_{n-2}\rangle$. It was found by Lanczos \cite{lanczos1950iteration} that once this is provided, $|K_n\rangle$ is also automatically orthogonal to all the other Krylov-basis states. Thus eq. \eqref{eq:KrylovBasis} implies that the Hamiltonian is tri-diagonal in the Krylov basis.

We work with a finite Hilbert space spanned by basis states $\{|v_i\rangle\}$ for $ i=0,\dots,\mathcal{D} -1$. Since we will consider quantum walks in this Hilbert space, the states $\{|v_i\rangle \}$ will be viewed as the vertices of a graph. A Hamiltonian $H$ on this space determines whether two vertices $|v_i\rangle$ and $|v_j\rangle$ are connected or not by having either of two values $\langle v_i | H|v_j\rangle = J$ or $0$, respectively (henceforth we measure energy in units of the coupling constant $J$, i.e. $J=1$). Since $\mathcal{C}(t)$ defined in eq. \eqref{eq:cmplxDef} is in general oscillatory, there is an inherent difficulty when comparing the complexities of two states. For this reason, we work with the long-time average of  the Krylov complexity that we derive next. This quantity will be used as a cost function in the maximization/minimization algorithm below. Starting from definition eq. \eqref{eq:cmplxDef}, insert resolutions of identity to obtain
\small
\begin{eqnarray}
\mathcal C(t)  = \sum_{n,m,l}  w_n \langle K_n | E_m\rangle  \langle E_m| K_0\rangle \langle K_0 |E_ l\rangle \langle E_l| K_n\rangle e^{i (E_l-E_m) t} 
\end{eqnarray}
\normalsize
where $|E_m\rangle$ and $|E_l\rangle$ are the exact eigenstates of $H$ with energies $E_m$ and $E_l$. 
Taking the long-time average of this expression yields
\begin{eqnarray}
\label{eq:avgC}
\bar{\mathcal C} &\equiv& \lim_{T\rightarrow \infty}\frac{1}{T}\int_{0}^{T} dt\  \mathcal C(t)  = \sum_{n} w_n \kappa_n \\
\kappa_n &\equiv&  \sum_m \left|\langle E_m| K_0\rangle \right|^2 \left|\langle K_n | E_m\rangle \right|^2  \label{eq:kappa}
\end{eqnarray}
This expression for the average $\bar{\mathcal C}$ was also introduced and studied in \cite{rabinovici2022krylov} and \cite{PhysRevLett.132.160402}. We assumed in eq. \eqref{eq:avgC} that $1/T \int_0^{T} e^{i(E_l-E_m)t} dt = \delta_{ml}$ when $T$ goes to infinity. This is valid provided that the spectrum is not degenerate. Since some of the Hamiltonians below turn out to be degenerate, we will combine the use of eq. \eqref{eq:avgC} with analytical calculations to ensure the correctness of the results.
\begin{figure}[t] 
\centering{}\includegraphics[width=\columnwidth]{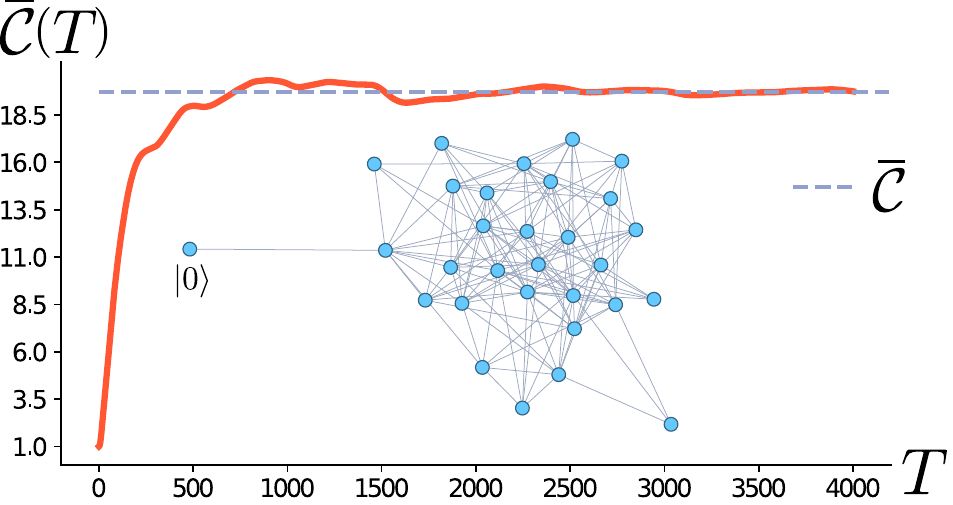}\caption{\label{Figconv} Shown is the time average $\bar{\mathcal{C}}(T) = \frac{1}{T}\int_{0}^{T} dt \  \mathcal C(t) $ as a function of $T$ for the displayed  graph with $\mathcal D = 30$ vertices.  As $T$ increases, this integral approaches the computed long-time average $\bar{\mathcal{C}}$  in eq. \eqref{eq:avgC}.}
\end{figure}
The typical convergence of the long-time average of $C(t)$ to $\bar{\mathcal C}$ is shown in Fig. \ref{Figconv}. We always seed the Krylov procedure with the node $|v_0\rangle$ of the graph, thus $|K_0\rangle=|v_0\rangle$. When we apply the graph optimization algorithm below we will obtain graphs that maximize or minimize the $\bar{\mathcal C}$ spread complexity of the state initialized in $|v_0\rangle$. These graphs will not in general extremize the  $\bar{\mathcal C}$ spread complexity of the other basis states.

\section{Relation to the quantum-walk limiting distribution}
In contrast to a classical random-walk, a quantum-walk never converges to a stationary state since the evolution is unitary rather than stochastic. Nevertheless, as shown by \cite{aharonov2001quantum} and later employed by \cite{childs2002example}, it is possible to introduce a quantity $\chi_i$, which is the long-time average of the probability to find the system in state $|v_i\rangle$. This is called the \emph{limiting distribution} of the quantum-walk. The authors of \cite{childs2002example} define $\chi_i$ formally as
\begin{eqnarray}
\chi_i &\equiv& \lim_{T \rightarrow \infty} \frac{1}{T} \int_0^T  |\langle v_i| e^{-i H t} |v_0\rangle |^2 dt \\
&=& \sum_{m} |\langle E_m | v_0 \rangle|^2 \ |\langle v_i  |E_m  \rangle|^2,
\end{eqnarray}
assuming for the last step that $H$ has no degeneracies. Clearly there is a resemblance of the $\chi$'s to the Krylov $\kappa$'s in eq. \eqref{eq:kappa}: The only difference is the appearance of  the Krylov-basis in place of the vertex-basis.  In fact, there are cases where the Krylov-basis and the vertex-basis coincide, like the path graphs of Fig. \ref{FigMain} that we also discuss in App. \ref{Sec:PathGraphs}. In such cases the $\chi$'s and $\kappa$'s are identical. In cases where only a few of the basis vectors coincide, we can still draw interesting conclusions, as we demonstrate next.

\begin{figure}[t] 
\centering{}\includegraphics[width=0.8\columnwidth]{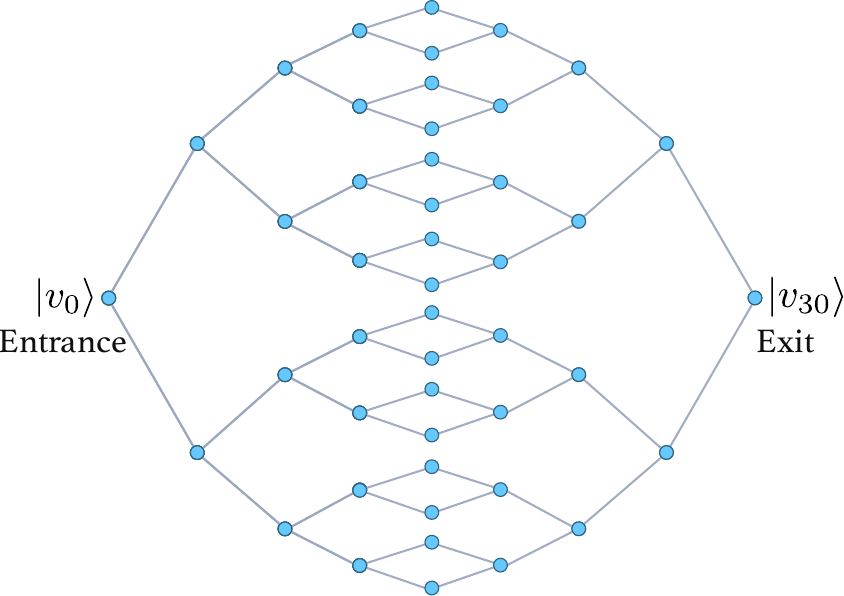}\caption{\label{GluedBinaryTree}  This graph structure was considered in \cite{childs2002example} to discuss the traversal time of the quantum walk. The system is initialized in $|v_0\rangle$ and time-evolved by the adjacency matrix Hamiltonian. The graph consists of two binary trees, one of height $n$, the other of height $n+1$ glued together at their leaves, here $n=4$.}
\end{figure}

Childs, Farhi and Gutman discussed the effectiveness of quantum walks in comparison to their classical counterparts \cite{childs2002example}. To illustrate the difference they considered a graph $G_n$ that consists of two binary trees, one of height $n$ and the other of height $n+1$, glued together at their leaves, see Fig. \ref{GluedBinaryTree} for an example of $G_4$. If the system is initialized in the root of the left binary tree, it finds the exit state at the root of the right tree in linear time. This is in stark contrast to the classical random walk where the walker takes random steps from node to neighboring node. When the walker arrives near the central column, it spends an exponentially (in $n$) long time there. The reason for this is that there are twice as many paths that lead into the central column as lead out of it. In \cite{childs2002example} it is shown that the efficiency of the quantum walk is tied to the fact that the $\chi$ value of the exit node is not exponentially small in $n$. The authors prove this by deriving the bound $\chi_{\text{Exit}} > 1/(2n+1)$. It turns out that the value of $\chi$ at the exit node is identical to the value of the Krylov $\kappa$ at the exit, since the last Krylov-basis vector is equal to the exit vertex-basis vector: $|K_{2n}\rangle = |v_{\text{Exit} }\rangle$. We show this in App. \ref{app:bintree} and also calculate its exact value:
\begin{eqnarray}
\chi_{\text{Exit}} = \kappa_{\text{Exit}}= \frac{3}{4n+4}
\end{eqnarray}
which is indeed larger than $1/(2n+1)$ for $n\geq 1$. We see that the Krylov $\kappa$ values themselves are good indicators for the efficiency of quantum walks in virtue of their connection to the $\chi$ values. Moreover, the Hilbert space dimension of the glued tree is $\mathcal D = 3 \cdot 2^n -1$, while the Krylov subspace has the exponentially smaller dimension $2n+1$, see App. \ref{app:bintree} for details. Then according to eq. \eqref{eq:avgC} the Krylov $\bar{\mathcal C}$-complexity cannot be larger than $w_{2n+1}$, in particular it cannot be as large as $w_{\mathcal D}$, which is exponentially larger. In this way, the $\bar{\mathcal C}$-complexity captures the ease with which the quantum walk traverses the glued tree from entrance to exit. 

\section{Complexity Extremization Algorithm and Analytically Tractable Graphs} To get a sense of the landscape of Krylov complexity values, we construct graph Hamiltonians that have extremal values of $\bar{C}$. Hamiltonians $H$ corresponding to such graphs have only two types of entries, \ $0$ or $1$, i.e. they are symmetric binary matrices. To avoid disconnected subspaces, we require that $H$ represent a connected graph, i.e. between any two vertices there should exist a path of vertices that joins them.

To generate the graphs of minimal and maximal $\mathcal{\bar{C}}$-complexity we employ a stochastic greedy algorithm. Then the cost function is either $-\mathcal{\bar{C}}$ or $+\mathcal{\bar{C}}$, depending on whether we want to maximize or minimize the complexity. The algorithm operates on a graph with $\mathcal D$ vertices, representing the vertex-basis states in Hilbert space. Then there are $2^{\mathcal D (\mathcal D-1)/2}$ possible graphs without self-loops that need to be considered. Thus a brute-force search in the space of possible Hamiltonians is out of the question even for small $\mathcal D$. Instead we proceed by a series of local optimizations. The algorithm selects a node $i$ at random and then selects up to $20$ neighbors of this node ${n_1,\dots,n_{20}}$ also at random. Then it considers all $2^{20}$  combinations of choosing each $H_{i,n_j} = H_{n_j,i}= 0 \text{ or } 1$ and selects the one that has the smallest $\bar{\mathcal{C}}$ value and also results in a connected graph. The latter condition is checked by running a depth-first-search from the node $|v_0\rangle$ and seeing if all of the nodes are visited in the process \cite{skiena1998algorithm}.   This optimizes the structure of the graph connected to node $i$. This sequence of steps is iterated until the cost function stops changing. Seeding the algorithm with different initializations, we observe that the program repeatedly finds the same lowest-cost graph. Occasionally, the algorithm gets stuck on graphs of higher cost and has to be restarted. The Krylov $\mathcal{\bar{C}}$-complexity itself is calculated using the expression in eq. \eqref{eq:avgC}, with $w_n=n$, and tri-diagonalizing the Hamiltonian by applying the Hessenberg decomposition as implemented in the GNU Scientific Library \cite{galassi1996gnu}. The author's C\texttt{++} implementation is available on GitHub \cite{githubcode}.
\begin{figure}[t] 
\centering{}\includegraphics[width=0.9\columnwidth]{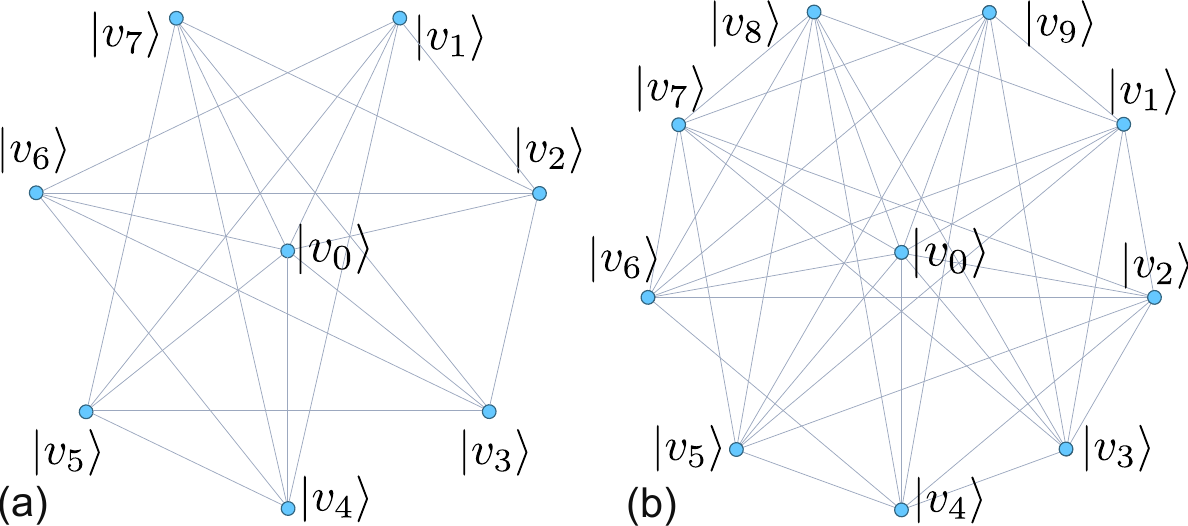}\caption{\label{FigLowCGraphs}  Graphs of low complexity produced by minimization algorithm. Both graphs have a hub vertex $|v_0\rangle$ connected to all the other vertices. If one were to remove this node and all its adjacent edges one would be left with a k-regular graph. The graph in (a) has $\mathcal D = 8$ vertices and $k=4$ while (b) has $\mathcal D = 10$ vertices and $k=6$. }
\end{figure}

Let us first discuss the Hamiltonians that give rise to the lowest $\bar{\mathcal{C}}$-complexity with the seed state $|v_0\rangle$. The algorithm is initialized from a symmetric random binary matrix. After several thousand iterations it finds certain graphs as the ones shown in Fig. \ref{FigLowCGraphs} . Closer inspection of all graphs reveals that the node $|v_0\rangle$, corresponding to the first basis state, is connected to all the other $\mathcal D -1$ vertices. In graph theory parlance $|v_0\rangle$ is a hub. What all the discovered graphs here have in common is that if one removes the hub and all its edges, the remainder of the graph is $k$-regular, i.e. each vertex has the same number $k$ of edges. It turns out that given this information we can analytically understand why the complexity of these graphs is so low. 

We begin by choosing for $H$ a graph that is $k$-regular and has in addition a hub $|v_0\rangle$ that is connected to all the other nodes. Then one can write down two exact eigenstates of such graphs. Any $k$-regular graph has the eigenvector $\sum_i |v_i\rangle$ with eigenvalue $k$. Since we have in addition to this a hub $|v_0\rangle$, we try an ansatz of the form 
\begin{eqnarray}
\label{eq:evecAnsatz}
|\psi_a\rangle \equiv a|v_0\rangle+\sum_{i=1}^{\mathcal D -1} |v_i\rangle
\end{eqnarray}
This turns out to yield two eigenvectors of $H$ as follows. First, read off from the eigenvector condition
\begin{eqnarray}
H \bm |\psi_a\rangle &=& (\mathcal D-1)|v_0\rangle+(k+a)\sum_{i=1}^{\mathcal D -1} |v_i\rangle \\
&=&\lambda \bm |\psi_a\rangle
\end{eqnarray}
the two relations
\begin{eqnarray}
a \lambda&=& \mathcal D-1 \\
k+a &=& \lambda.
\end{eqnarray}
It turns out that the equations can be satisfied in two ways:
\begin{eqnarray}
a_\pm&=&-\frac{k}{2} \pm\frac{\sqrt{4 (\mathcal D -1)+k^2}}{2}\\
\lambda_\pm &=& \frac{k}{2} \pm\frac{\sqrt{4 (\mathcal D-1)+k^2}}{2}
\end{eqnarray}
Thus we have found two eigenstates of $H$ of the form \eqref{eq:evecAnsatz}.We denote these by  $|\psi_\pm\rangle$ and their normalized versions by $|\pm\rangle$.

Next we note that according to eq. \eqref{eq:evecAnsatz} the difference between $|\psi_+\rangle$ and $|\psi_-\rangle$ is parallel to the first Krylov vector $|K_0\rangle=|v_0\rangle$. Hence any power of $H$ acting on $|K_0\rangle$ lies in the  two-dimensional subspace spanned by $|\psi_\pm\rangle$. Consequently the Krylov subspace is only two-dimensional with basis vectors
\begin{eqnarray}
|K_0\rangle &=& \alpha |+\rangle + \beta |-\rangle \\
|K_1\rangle &=& -  \beta |+\rangle +\alpha |-\rangle
\end{eqnarray}
where $\alpha, \beta$ account for normalization and $\alpha^2 + \beta^2 = 1$.
Thus the spread-complexity defined in eq. \eqref{eq:cmplxDef} has only one non-zero term:
\begin{eqnarray}
\mathcal C(t) &=& |\langle K_1|\psi(t)\rangle|^2 \\
&=& \frac{4\mathcal D - 4}{4\mathcal D - 4+k^2} \sin^2 \left( \frac{\lambda_+ - \lambda_-}{2}t\right) ,
\end{eqnarray}
yielding the $\bar{\mathcal C}$-complexity
\begin{eqnarray}
\bar{\mathcal C} =2\frac{\mathcal D - 1}{4\mathcal D - 4+k^2} .
\end{eqnarray}
The first point to notice is that the complexity $C(t)$ does not distinguish between the specific kind of attached $k$-regular graph: all the different $k$-regular graphs with appended hub have the same spread-complexity.  Secondly, the complexity of this class of graphs is minimized by choosing $k$ maximally, i.e. by setting $k=\mathcal D-2$. This yields the lowest possible non-zero $\bar{\mathcal C}$-complexity value
\begin{eqnarray}
\bar{\mathcal C}_\text{min} =2\frac{\mathcal D - 1}{\mathcal D^2} .
\end{eqnarray}
Together with the hub and its edges, this is the \emph{complete graph} on $\mathcal D$ vertices, see Fig. \ref{FigMain}. We note that in the thermodynamic limit $\mathcal D \rightarrow \infty$, the $\bar{\mathcal C}$-complexity of the complete graph goes to $0$ as $\bar{\mathcal C}_\text{min} \sim 2/\mathcal D$. 

The star graph in Fig. \ref{FigMain} also falls into the previous class of graphs by choosing $k=0$. Thus its complexity is 
\begin{eqnarray}
\bar{\mathcal C}_\star =\frac{1}{2}
\end{eqnarray}
irrespective of the number of vertices $\mathcal D$.

We next turn to the graphs with maximum $\bar{\mathcal{C}}$-complexity for seed state $|v_0\rangle$. From eq. \eqref{eq:cmplxDef}, it is clear that $\bar{\mathcal{C}}_{\max}\leq \mathcal D$. We find that this bound is never attained. Examples of graphs found by the algorithm are shown in Fig. \ref{FigMain} (blue squares) for up to $30$ nodes. All these graphs have in common that the seed vertex $|v_0\rangle$ is a node of degree $1$ (a leaf). In contrast to the case of minimum $\bar{\mathcal{C}}$-complexity, here the graph with maximum $\bar{\mathcal{C}} $-complexity is unique. The data shown in Fig. \ref{FigMain} suggests that the maximum possible $\bar{\mathcal{C}}$-complexity for graphs scales linearly with the Hilbert space dimension, a linear fit results in the empirical rule
\begin{eqnarray}
\label{maxComplxty}
\bar{\mathcal{C}}_{\max} =  0.66 \mathcal{D} -1.31,
\end{eqnarray}
for large $\mathcal D$. Thus all the graphs we studied have $\bar{\mathcal{C}}$  complexities that lie in the shaded wedge in Fig. \ref{FigMain}.  We can exhibit another class of graphs that also have linear $\bar{\mathcal{C}}$-complexity scaling. Consider the class of \emph{path graphs}, i.e. a linear arrangement of vertices where neighbors are connected, see  Fig. \ref{FigMain}. It is shown in App. \ref{Sec:PathGraphs} that the Krylov $\bar{\mathcal{C}}$-complexity for this class of graphs is 
\begin{eqnarray}
\bar{\mathcal C}_\text{Path Graph} = \frac{1}{2}\mathcal D -\frac{1}{2}
\end{eqnarray}
thus their $\bar{\mathcal{C}}$ complexity also scales linearly, see the dotted line in Fig. \ref{FigMain}. Typically the Hilbert space dimension of physical systems scales exponentially with the degrees of freedom. Thus the last two examples of graphs have $\bar{\mathcal{C}}$-complexities that grow exponentially with system size.
\section{Outlook}
In summary, we have investigated the Krylov spread-complexity in the context of quantum walks and established a connection with the well-known limiting-distribution. We found that the $\bar{\mathcal{C}}$-complexity of graphs can be very different functions of the Hilbert space dimension: complete graphs have $\bar{\mathcal{C}}$-complexities that decrease as the inverse of the dimension, while the $\bar{\mathcal{C}}$ of path-graphs and the maximum graphs grow linearly. Between these lie the complete trees with only logarithmically growing $\bar{\mathcal{C}}$. 

It remains an open question whether the Krylov $\bar{\mathcal{C}}$-complexity can also be used as a practical tool to assess the difficulty of state-preparation tasks \cite{araujo2021divide,10.5555/2011670.2011675, zhang2021low}. With the advent of a large number of NISQ platforms, graphs have played a prominent role and have been realized with Gaussian boson samplers \cite{deng2023solving}, superconducting processors \cite{harrigan2021quantum}, integrated photonic devices \cite{caruso2016fast} and Rydberg atom arrays \cite{byun2022finding, jeong2023quantum,ebadi2022quantum}. Therefore, these platforms are ideal for future explorations of Krylov $\bar{\mathcal{C}}$-complexity and to test its utility as a proxy for gate-complexity.

\appendix
\begin{widetext}
\section{Krylov Complexity of the Path Graph}\label{Sec:PathGraphs}
We consider the path graph on $\mathcal D$ vertices, shown in Fig. \ref{Fig:PathGraph}.
\begin{figure}[b]
\centering{}\includegraphics[width=0.4\textwidth]{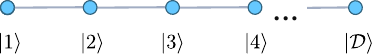}\caption{\label{Fig:PathGraph} The path graph on $\mathcal D$ vertices.}
\end{figure}
The adjacency matrix $H$ corresponds, of course, to a tight-binding model for a particle that is hopping on a line with open boundary conditions. The $m$-th eigenvector of the adjacency matrix has the eigenvalue 
\begin{eqnarray}
E_m = 2\cos\left(\frac{\pi}{\mathcal D+1}m\right)
\end{eqnarray}
for $m=1,\dots,\mathcal D$ and eigenvector
\begin{equation}
\psi_{m}=\sqrt{\frac{2}{\mathcal D +1}}\left(\sin k ,\sin 2 k ,\dots,\sin\mathcal D k \right)^{T} 
\end{equation}
with $k \equiv \frac{\pi m}{\mathcal D+1}$.  Since the Hamiltonian is already in tri-diagonal form, the vertex-basis states $\{|i\rangle \}$ are also the Krylov-basis states generated from state $|0\rangle$. Then according to eq. \eqref{eq:kappa} we have for the $\kappa$ values
\begin{eqnarray}
\label{eq:pathGraphkappa}
\kappa_n &=& \frac{4}{(\mathcal D+1)^2}\sum_{m=1}^{\mathcal D}\sin^2\left(\frac{\pi m(n+1)}{\mathcal D +1 } \right)\sin^2\left(\frac{\pi m}{\mathcal D +1 } \right)= 
\begin{cases}
\frac{3}{2\mathcal{D}+2} & \text{if }n=0\text{ or } \mathcal{D}-1\\
\frac{1}{\mathcal{D}+1} & \text{otherwise.}
\end{cases}
\end{eqnarray}
and consequently for the long-time average 
\begin{eqnarray}
\label{eq:pathGraphC}
\bar{\mathcal C}_\text{Path Graph} &=& \sum_{n=0}^{\mathcal D -1}\kappa_n n=\frac{\mathcal D-1}{2}.
\end{eqnarray}
\section{Krylov Complexity of complete m-ary Trees and Childs-Farhi-Gutmann Trees $G_n$}\label{app:bintree}
In the case of binary trees, and in fact all n-ary trees, the computation of the Krylov-basis vectors is straightforward.
\begin{figure}[b]
\centering{}\includegraphics[width=0.4\columnwidth]{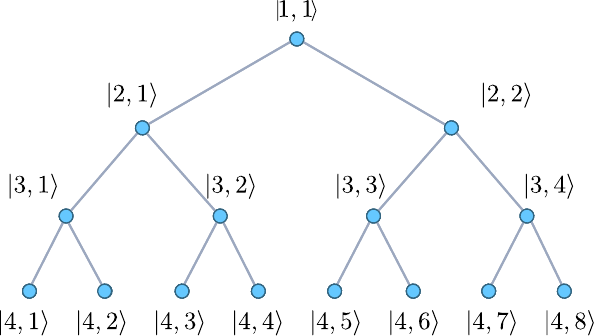}\caption{\label{FigBinTree} Labeling convention for the binary tree.}
\end{figure}
 We illustrate the calculations first in terms of binary trees. We label each vertex of the binary tree as $|i,j\rangle$ where $i$ is the height of the node and $j$ the horizontal position, see Fig. \ref{FigBinTree}.  The Krylov-basis construction begins with the root node 
 \begin{eqnarray}
 |K_0\rangle \equiv |1,1\rangle.
 \end{eqnarray}
 Denoting the adjacency matrix of the tree by $H$, we note that $H$ acting on $|K_0\rangle$ produces $|2,1\rangle$+$|2,2\rangle$. Thus 
 \begin{eqnarray}
 |K_1\rangle \equiv \frac{1}{\sqrt{2}}\left[ |2,1\rangle+|2,2\rangle\right].
 \end{eqnarray}
 When $H$ acts on $|K_1\rangle$ it produces the state 
 \begin{eqnarray}
H |K_1\rangle =  \frac{1}{\sqrt{2}} |1,1\rangle+\frac{1}{\sqrt{2}}\left[ |3,1\rangle+|3,2\rangle+|3,3\rangle+|3,4\rangle\right].
 \end{eqnarray}
 We can immediately orthonormalize this state with respect to the previous Krylov-vectors by subtracting the first term and normalizing the remaining part:
 \begin{eqnarray}
|K_2\rangle \equiv  \frac{1}{2}\left[ |3,1\rangle+|3,2\rangle+|3,3\rangle+|3,4\rangle\right],
 \end{eqnarray}
 which is an equal-weight superposition of all height $3$ states. Continuing like this, we find that the next Krylov vector is an equal-weight superposition of all height $4$ states
 \begin{eqnarray}
|K_3\rangle \equiv  \frac{1}{\sqrt{2^3}}\left[ |4,1\rangle+|4,2\rangle+|4,3\rangle+|4,4\rangle+|4,5\rangle+|4,6\rangle+|4,7\rangle+|4,8\rangle\right].
 \end{eqnarray}
 Further action of $H$ on these vectors produces no new basis vectors. Thus the Hamiltonian on this Krylov subspace has the simple form
  \begin{eqnarray}
H_{1,1} \equiv \sqrt{2}\left(\begin{array}{cccc}
0 & 1 & 0 & 0\\
1 & 0 & 1 & 0\\
0 & 1 & 0 & 1\\
0 & 0 & 1 & 0
\end{array}\right),
 \end{eqnarray}
 where the indices of $H$ indicate that this is the Krylov subspace spawned by the state $|1,1\rangle$. 
 For a general binary tree of height $h$  the Krylov vectors are also equal-weight superpositions of all nodes with identical heights
 \begin{eqnarray}
 \label{eq:KrylovBinTree}
|K_i \rangle &\equiv&  \frac{1}{\sqrt{2^{i}}}\sum_{j=1}^{2^{i}}|i+1,j\rangle \\
 \end{eqnarray}
 for $i=0,\dots,h-1$. Then the Hamiltonian on this subspace is the $h\times h$ matrix
  \begin{eqnarray}
  \label{eq:KrylovSubspaceH}
H_{1,1} \equiv \sqrt{2}\left(\begin{array}{cccccc}
0 & 1 & 0 & 0 & 0 & 0\\
1 & 0 & 1 & 0 & 0 & 0\\
0 & 1 & 0 & 1 & 0 & 0\\
0 & 0 & 1 & 0 & \ddots & 0\\
0 & 0 & 0 & \ddots & \ddots & 1\\
0 & 0 & 0 & 0 & 1 & 0.
\end{array}\right)
 \end{eqnarray}
 This is once more the tight-binding model. We see that the binary tree in terms of Krylov vectors is transformed into a path graph of length $h$. Thus we can read off the eigenvalues and eigenvectors:
 \begin{eqnarray}
E_m = 2\sqrt{2}\cos\left(\frac{\pi}{h+1}m\right)
\end{eqnarray}
with eigenvector
\begin{equation}
\psi_{m}=\sqrt{\frac{2}{h +1}}\left(\sin k ,\sin 2 k ,\dots,\sin h k \right)^{T} ,
\end{equation}
where $k \equiv \frac{\pi m}{h+1}$. The fundamental change compared to the path graph is the fact that the Hilbert space dimension $\mathcal D$ only enters through the height $h$ of the tree. But since $h=\log_2{(\mathcal D +1)}$ the Krylov $\bar{\mathcal C}$-complexity is now logarithmically lower. In fact it is equal to
\begin{eqnarray}
\bar{\mathcal C}_\text{Binary Tree} &=& \frac{h-1}{2} \sim \frac{1}{2} \frac{\log \mathcal D}{\log 2},
\end{eqnarray}
which is much less than for a path graph of the same Hilbert space dimension.

We can now use these results to calculate the $\bar{\mathcal C}$-complexity of the glued binary tree considered by Childs, Farhi and Gutmann \cite{childs2002example}, see Fig. \ref{GluedBinaryTree}. The glued tree of order $n$, denoted by  $G_n$, is obtained by gluing together a binary tree of height $n$ to another binary tree of height $n+1$. First note that by the same reasoning as before, the Krylov-basis vectors are still the equal-weight superpositions of nodes of equal height. Secondly, the total number of Krylov-basis vectors is equal to $n+n+1=2n+1$, one for each level of the graph. The total number of nodes, i.e. the Hilbert space dimension,  $\mathcal D = 2^n-1 +2^{n+1}-1=3\cdot 2^n-1$ Then the Hamiltonian in this Krylov subspace has dimensions $(2n+1)\times(2n+1)$ and has the same form as eq. \eqref{eq:KrylovSubspaceH}, thus it is 
also a path graph. The $\kappa$ values and $\bar{\mathcal C}$-complexity for the glued binary trees $G_n$ are then simply obtained from eqs. \eqref{eq:pathGraphkappa} and \eqref{eq:pathGraphC} as
\begin{eqnarray}
\kappa_l &=& \ 
\begin{cases}
\frac{3}{4n+4} & \text{if }l=0\text{ or } 2n\\
\frac{1}{2n+2} & \text{otherwise.}
\end{cases}\\
\bar{\mathcal C}_{G_n} &=& n \sim \frac{\log \mathcal D/ 3}{\log  2}. 
\end{eqnarray}

Finally, we comment briefly on the $\bar{\mathcal C}$-complexity of the complete $m$-ary tree. This is defined as a tree for which all nodes, with the exception of the leaves, have exactly $m$ children nodes. We can repeat the Krylov-basis construction from before and obtain a Krylov-vector for each level of the tree that is an equal-weight superposition of all the nodes at that level. The total number of nodes is equal to the Hilbert space dimension $\mathcal D$. A complete $m$-ary tree of height $h$ has a total of $\mathcal D = (m^{h+1}-1)/(m-1)$ nodes. Thus we see that for large $\mathcal D$ we have $h\sim \log D/\log m$ and 
\begin{eqnarray}
\bar{\mathcal C}_{m\text{-ary tree} } &\sim& \frac{1}{2}\frac{\log{\mathcal D}}{\log m}. 
\end{eqnarray}
\end{widetext}
\bibliography{biblio}
\end{document}